# Estimation of radiative half-life of $^{229m}$Th by half-life measurement of other nuclear excited states in $^{229}$Th

Y. Shigekawa,[1,*] A. Yamaguchi,[2,3] K. Suzuki,[4] H. Haba,[1] T. Hiraki,[4] H. Kikunaga,[5] T. Masuda,[4] S. Nishimura,[1] N. Sasao,[4] A. Yoshimi,[4] and K. Yoshimura[4]

[1]Nishina Center for Accelerator-Based Science, RIKEN, Wako, Saitama 351-0198, Japan

[2]Quantum Metrology Laboratory, RIKEN, Wako, Saitama 351-0198, Japan

[3]PRESTO, Japan Science and Technology Agency, Kawaguchi-shi, Saitama 332-0012, Japan

[4]Research Institute for Interdisciplinary Science, Okayama University, Okayama, Okayama 700-8530, Japan

[5]Research Center for Electron Photon Science, Tohoku University, Sendai, Miyagi 982-0826, Japan

**Abstract**

We perform coincidence measurements between α particles and γ rays from a $^{233}$U source to determine the half-lives of the excited state in a $^{229}$Th nucleus. We first prove that the half-lives of 42.43- and 164.53-keV states are consistent with literature values, whereas that of the 97.14-keV state (93(7) ps) deviates from a previously measured value (147(12) ps). The half-lives of 71.83- and 163.25-keV states are determined for the first time. Based on the obtained half-lives and the Alaga rule, we estimate the radiative half-life of the low-energy isomeric state ($^{229m}$Th) to be 5.0(11)×10$^3$ s, which is one of the key parameters for the frequency standard based on $^{229}$Th.

*yudai.shigekawa@riken.jp

## I. Introduction

The first excited state in the $^{229}$Th nucleus ($^{229m}$Th) has an extremely low excitation energy of ~8 eV [1–4]. The low-lying isomeric state allows the laser excitation and spectroscopy of the nucleus, potentially leading to an ultraprecise optical nuclear clock [5,6]. Recent experimental studies confirmed the existence of $^{229m}$Th and its decay by the internal conversion (IC) process [7–9], and constrained the excitation energy of $^{229m}$Th accurately enough to initiate laser excitation experiments [1–4].

The radiative half-life of $^{229m}$Th determines the natural linewidth of the nuclear transition between its ground and isomeric states, which is used as a reference transition of the $^{229}$Th nuclear clock. Hence, determination of the radiative half-life of $^{229m}$Th is important to estimate the clock performance. The radiative half-life of $^{229m}$Th has been theoretically estimated to be $10^3$–$10^4$ s [10–12] hitherto, and its lower limit was experimentally measured to be 60 s [7]. The radiative half-life of $^{229m}$Th can be also obtained from the reduced transition probability $B(X\lambda)$ of the isomeric decay that is estimated from the $B(X\lambda)$ values of the inter-rotational-band transitions beginning with higher excited states than $^{229m}$Th by applying the Alaga rule [10,13]. The Alaga rule states that the ratio of two $B(X\lambda)$ values for a pair of intra- or inter-band transitions equals the ratio of the squares of Clebsch-Gordan coefficients based on the assumption of the separable rotational motion of a nucleus [14].

To obtain $B(X\lambda)$ values and apply the Alaga rule, decay properties (e.g. γ branching ratios and half-lives) and nuclear properties (e.g. nuclear spins and energies) should be known. Such properties for the $^{229}$Th nucleus have been investigated via the spectroscopy of γ rays following the α decay of $^{233}$U [1,3,15–20], β decay of $^{229}$Ac [11,21,22], and electron capture decay of $^{229}$Pa [23]. Furthermore, α-particle spectroscopy of $^{233}$U [24,25], IC-electron spectroscopy [22,26,27], particle spectroscopy in the $^{230}$Th($d,t$)$^{229}$Th reaction [28,29], and a Coulomb excitation study of $^{229}$Th [30] have been conducted. These spectroscopic studies pertaining to transitions in the $^{229}$Th nucleus have identified more than 50 excited states and their energies, spins, and parities were determined, as summarized in [31]. The studies also determined the γ branching ratios, IC coefficients, and mixing ratios for some transitions. The levels related to this study are shown in Fig. 1.

The half-life measurement of excited states in the $^{229}$Th nucleus has been reported in a few previous studies. Ton *et al*. [32] performed a coincidence measurement between α-particles and IC electrons from a $^{233}$U source and reported the half-lives of the 42.43-keV ($I^{\pi} = 7/2^+$) and 97.14-keV ($I^{\pi} = 9/2^+$) states, which belong to rotational band 5/2$^+$[633], whose band head is the ground state of $^{229}$Th. Gulda *et al*. [22] and Ruchowska *et al*. [11] determined the half-lives of some excited states that belong to 5/2$^-$[752], 3/2$^-$[761], 1/2$^+$[631], 7/2$^+$[613], 5/2$^+$[622], and 1/2$^-$[770] bands via coincidence measurements of β$^-$ particles and subsequent γ rays from $^{229}$Ac. Recently, the half-life of the 29.19-keV state in the 3/2$^+$[631] band was determined by populating the state from the ground state with synchrotron radiation [33].

To estimate the $B(X\lambda)$ value of the decay of $^{229m}$Th based on the Alaga rule, the $B(X\lambda)$ values of the inter-band transitions between the 5/2$^+$[633] and 3/2$^+$[631] rotational bands beginning with the higher excited states than $^{229m}$Th are required. However, such $B(X\lambda)$ values have been experimentally obtained only for the 97.14 keV→71.83 keV transition [11,17,22,30], and the reported values were inconsistent among different studies [13]. The determination of half-lives for states belonging to the 5/2$^+$[633] and 3/2$^+$[631] bands, such as the 71.83-keV state, would improve the estimation of the $B(X\lambda)$ value and radiative half-life of the decay of $^{229m}$Th.

In this paper, we report the half-life measurement of the excited states in $^{229}$Th (primarily the states in the 5/2$^+$[633] and 3/2$^+$[631] bands) via coincidence measurements between γ rays and α particles emitted from a $^{233}$U source. The γ-ray measurement using a LaBr$_3$(Ce) scintillator performed in this study affords a higher energy resolution than the IC-electron measurement using a plastic scintillator [32]. Aided also by high-resolution α-particle measurement realized by a thin $^{233}$U source, the γ-ray transitions used for the half-life determination can be unambiguously selected. We reproduced the half-lives of the 42.43- and 164.53-keV states, whereas we found the half-life of the 97.14-keV state measured in this study was shorter than the value reported in [32]. For the first time, we determined the half-lives of the 71.83- and 163.25-keV states belonging to the $3/2^+$[631] and $5/2^+$[633] bands, respectively. We obtained the $B(X\lambda)$ values from the half-lives, and then estimated the radiative half-life of $^{229m}$Th based on the Alaga rule.

## II. Experimental procedures

### A. Preparation of $^{233}$U source

The $^{233}$U sample used in this study was chemically purified to remove the daughters of $^{233}$U and $^{232}$U via anion-exchange chromatography as described in [34]. The contamination of $^{232}$U in the $^{233}$U sample was measured to be 0.23 ppm by mass. The $^{233}$U stock solution ($^{233}$U 1.8 μg/μL in 0.01 M HNO$_3$) was subjected to electrodeposition [35] on an Al foil (18 μm thick) as follows: First, 20 μL of 0.01 M HNO$_3$ without $^{233}$U and 6.0 mL of 2-propanol were poured into an electrodeposition cell (inner diameter 24 mm) placed on the Al foil. A voltage of +400 V was applied to a Pt anode soaked in the solution for 20 min, and then the solution was removed from the cell. Next, we poured 20 μL of the $^{233}$U stock solution and 6.0 mL of 2-propanol into the cell and performed the electrodeposition of $^{233}$U by applying +200 V to the Pt anode for 30 min. The current during electrodeposition was 1.1–1.3 mA. After electrodeposition, the solution was removed from the cell, and the $^{233}$U source on the Al foil was dried using an infrared lamp.

After the electrodeposition, the $^{233}$U source was subjected to α-particle spectrometry involving a Si PIN photodiode detector (Hamamatsu S3204-09), from which the radioactivity of $^{233}$U in the source (diameter 24 mm) was evaluated to be 12.9(2) kBq (electrodeposition yield ~100%). The $^{233}$U peaks in the spectrum indicated a full width at half maximum (FWHM) of 23 keV at 4824 keV, which was similar to the intrinsic energy resolution of the Si detector (~18 keV) used in this study. This indicates that a sufficiently thin $^{233}$U source was prepared for the energy selection of α particles in the coincidence measurement.

## B. Coincidence measurement

Fig. 2 shows the setup of the coincidence measurement. We placed the $^{233}$U source and a passivated implanted planar Si (PIPS) detector (Canberra TMPD450-20-300AM, timing resolution 200 ps in the FWHM), which was used for the α-particle measurement, inside a vacuum chamber. For the γ-ray measurement, a LaBr$_3$(Ce) scintillator (25.4 mm in diameter, 25.4 mm thick) and a photomultiplier (PMT, Hamamatsu R13449-100-10) were placed near a Kapton® window (0.1 mm thick) outside the vacuum chamber. The timing resolution of the LaBr$_3$(Ce) + PMT system was measured to be 360 ps at the FWHM. The loss of low-energy γ rays emitted from the $^{233}$U source due to absorption were minimized through the thin Al backing foil of $^{233}$U and the thin Kapton® window. The vacuum chamber was enclosed by 5-cm Pb blocks to reduce background γ rays and was evacuated using a rotary pump.

Fig. 3 shows the block diagram of the coincidence measurement. The charges from the PIPS detector were collected using a charge-sensitive preamplifier (Tennelec TC 170), which yielded timing and energy signals. The signals from the PMT were divided into timing and energy signals. For both detectors, the timing signals were amplified, discriminated, divided, and sent to a coincidence module and a time-to-digital converter (TDC, REPIC RPC-060). The energy signals from the PIPS detector were amplified using a pulse-shaping amplifier (ORTEC 572A), and the pulse heights were recorded using a peak-hold analog-to-digital converter (ADC, Phillips Scientific 7164). After the energy signals from the PMT were amplified, the total charges were recorded using a charge-integrating ADC (Hoshin electronics C009). Each list-data recording using the computer-aided measurement and control (CAMAC) modules including the TDC and ADC modules was activated by the gate created by the coincidence module at the time of coincidence of the timing signals from both detectors.

We performed 791 sets of measurements, each of which comprised a 1-h coincidence measurement and a 1-min ADC pedestal measurement. The entire measurement lasted approximately 35 days. The TDC values were calibrated by a pulse delay generator whose timebase was a GPS-disciplined clock. The calibration error was negligibly small compared with errors of half-lives reported in this study.

### C. Data analysis

To determine the half-lives, time traces of γ-ray signals following the α-particle signals were constructed using the TDC values for the PIPS detector and PMT as follows: First, for each dataset fixed for 1 h, we constructed a histogram of the TDC values for each detector. Subsequently, a Gaussian function was fitted to the histogram to obtain the mean TDC value for each dataset. The mean TDC value drifted as time progressed and was therefore subtracted from the TDC value of each event. The corrected TDC value of each event for the PIPS detector was subtracted from that for the PMT to obtain the time trace.

The two-dimensional (2D) energy spectrum of the coincided α-particle and γ-ray events (Fig. 4) used to determine the energy windows of the time traces was constructed as follows: The ADC values for both detectors were first subjected to pedestal correction. Subsequently, the histograms of the ADC values for both detectors were constructed for each dataset for 1 h. The strongest peak in each histogram for the PIPS detector and PMT was the 4783.5-keV α peak and 13.0-keV X-ray peak, respectively. The mean values of these peaks obtained via Gaussian fitting drifted as time progressed. Hence, all of the ADC values were corrected such that the mean value for each dataset would become that for the first dataset. By combining the corrected ADC values for all the datasets, the 2D energy spectrum was obtained. Energy calibration was performed using the observed peaks. The FWHMs for the 4783.5-keV α peak (strongest peak) and 42.43-keV γ peak (second strongest peak) were 26 and 7 keV, respectively.

## III. Results

### A. Half-lives of 42.43- and 164.53-keV states

Fig. 5(b) shows the γ-ray spectrum gated by the 4783.5-keV α particles feeding the 42.43-keV state (Fig. 5(a)). The high-energy side of the α energy window was 2σ (σ: standard deviation) from the center of the 4783.5-keV α peak, whereas 0.77σ was selected for the low-energy side of the α energy window to avoid interference from the α particles feeding energy states higher than the 42.43-keV state, as confirmed in Fig. 5(b) by the negligibly small counts of γ rays higher than 50 keV. Fig. 5(c) shows the time trace of the 42.43-keV γ rays following the 4783.5-keV α particles. A single exponential decay function convoluted with a Gaussian function was fitted to the time trace using the maximum-likelihood method to obtain the half-life and its uncertainty. Here fitting parameters were the half-life in the single exponential decay function and the intensity, mean, and FWHM in the Gaussian function. The constant background was not included in the fitting function because the constant background counts (accidental coincidence counts) are negligibly small. The half-life of the 42.43-keV state was determined to be 169(4) ps. This half-life is consistent with 172(6) ps reported by Ton *et al* [32].

Similar to the 42.43-keV state, the half-life of the 164.53-keV state belonging to the 3/2$^-$[761] band was determined to be 70(50) ps with energy gates of the 4662.5-keV α and 164.52-keV γ peaks, as shown in Fig. 6. This half-life is consistent with 61(7) ps reported in [11].

### B. Half-life of 97.14-keV state

For the half-life determination of the 97.14-keV state, the γ-ray spectrum was gated by the appropriate energy window of the 4729-keV α particles, where γ rays with energies higher than 110 keV almost disappeared (Fig. 7(b)). The fitting of a single exponential decay function convoluted with a Gaussian function to the time trace for the 54.70-keV γ rays (Fig. 7(c)), corresponding to the 97.14 keV→42.43 keV transition, yielded a half-life of 103(12) ps. Moreover, the half-life can be obtained by selecting the 97.14-keV γ rays corresponding to the 97.14 keV→0 keV transition. In this case, the half-life was measured to be 88(9) ps (Fig. 7(d)). Both results agreed with each other within errors, and the weighted average of these values yielded a half-life of 93(7) ps. This half-life is inconsistent with 147(12) ps [32], which was obtained from the measurement of IC electrons following an α-particle emission from $^{233}$U using a plastic scintillator, which typically provides poor energy resolution.

**C.  Half-life of 71.83-keV state**

Fig. 8 (b) shows the α-particle spectrum gated by the 71.82-keV γ peak, where the 4754-keV α peak was distinguished from the 4729-keV peak owing to the gating. A single exponential decay function convoluted with a Gaussian function was fitted to the corresponding time trace (Fig. 8 (d)), yielding a half-life of 120(40) ps for the 71.83-keV state.

Because the 4754-keV α peak was not completely separated from the 4729-keV α peak, the γ-ray spectrum gated by the 4754-keV α peak (Fig. 8 (c)) showed a small peak in the 97.14-keV γ region, indicating that the population of the 97.14-keV states occurred slightly. As shown in Fig. 8 (a), this resulted in the emission of 67.95-keV γ rays (97.14 keV→29.19 keV), which could not be discriminated from the 71.82-keV γ peaks, or the inter-band population of the 71.83-keV state from the 97.14-keV state, possibly reducing the accuracy of the half-life. However, the γ rays observed in the 71.82-keV region originating from the population of the 97.14-keV state are only ~3% of those originating from the direct population of the 71.83-keV state, based on the counts of the 97.14-keV γ peak in Fig. 8 (c) and the branching ratios from the 97.14-keV state [31]. This value is almost negligible compared with the statistical error of the half-life (35%). Furthermore, we confirmed that the half-lives did not exhibit systematic variation when we slightly shifted the α energy window from the window used to construct the time spectrum in Fig. 8 (d) (4750–4780 keV).

## D. Half-life of 163.25-keV state

Fig. 9(b) shows the γ-ray spectrum gated by the 4664-keV α particles feeding the 163.25-keV state. A peak appears at ~66 keV, which could not be separated without gating. This peak corresponds to the 66.12-keV γ-rays from the 163.25 keV→97.14 keV transition (Fig. 9(a)). The time trace for a γ energy window of 64–71 keV (Fig. 9(c)) yielded a half-life of 220(30) ps for the 163.25-keV state by fitting a single exponential decay function convoluted with a Gaussian function.

As shown in Fig. 9(a), the 163.25-keV state decays into the 97.14- and 71.83-keV states via several paths, and the γ rays of 67.95 keV (97.14 keV→ 29.19 keV) and 71.8 keV (71.83 keV→0.008 or 0 keV) interfere with the 66.12-keV peak. The intensities of the 67.95 and 71.8 keV γ rays are calculated to be 0.6% and 7.2% of the intensity of the 66.12-keV peak, respectively [31]; the value for the 67.95-keV γ ray is negligibly small. The interference from the 71.8-keV γ peak with the 66.12-keV γ peak was reduced by selecting an appropriate γ energy window; we observed no clear systematic effects on the half-life when we narrowed the high-energy side of the γ energy window from 71 to 68.5 keV (Fig. 9(d)).

Meanwhile, the α particles of 4664 keV feeding the 163.25-keV state are not completely distinguished from those of 4681, 4678.6, and 4656 keV (Fig. 9(a)), which correspond to the populations of the 146.36-, 148.17-, and 173.48-keV states in the 5/2$^-$[752] band, respectively. These states decay to the 71.83-keV state and the 71.8-keV γ rays are emitted. Considering the α branching ratios to these states and the γ branching ratios from these states [31], the total probability of the 71.8-keV γ rays via the 146.36-, 148.17-, and 173.48-keV states being emitted is $1.9 \times 10^{-7}$ per α decay of $^{233}$U, which is only 2.8% of the probability of the 66.12-keV γ-rays being emitted via the 163.25-keV state. Hence, the interference from the 71.8-keV γ rays is almost negligible.

Another possible interference is the decay of the 146.36- and 148.17-keV states into the 71.83-keV state, which causes the emission of the 74.54- and 76.35-keV γ rays, respectively. The total probabilities of the 74.54- and 76.35-keV γ rays being emitted are calculated to be $1.1 \times 10^{-5}$ and 7.4×10$^{-7}$ per α decay of $^{233}$U, respectively [31]. The value for the 74.54-keV γ ray is comparable to the total probability of the 66.12-keV γ rays being emitted (6.7×10$^{-6}$ per α decay of $^{233}$U [31]). To observe the interference from the 74.54-keV γ rays, we compared the half-life obtained by fitting as a function of the high-energy side of the γ energy window, as shown in Fig. 9(d). The half-life obtained with the widest energy window (64–80 keV) was 350(22) ps, which was consistent with the half-life of the 146.36-keV state (329(8) ps [11]); this implies that the interference from the 146.36-keV state occurred in the energy window. Decreasing the high-energy side of the energy window from 80 keV reduced the half-life, and the half-life became constant at 71 keV, as shown by the arrow in Fig. 9(d), where the interference from the 146.36-keV state became almost negligible.

## IV. Discussion

The radiative half-life of $^{229m}$Th $T_{1/2,\gamma,\text{isomer}}$ is derived as follows:

$$T_{1/2,\gamma,\text{isomer}} = 3.94 \times 10^4 E_{\text{isomer}}^{-3} B(\text{M1}; 8.28 \text{ eV} \to 0 \text{ eV})^{-1} \text{ s}, \quad (1)$$

where $E_{\text{isomer}}$ is the excitation energy of $^{229m}$Th in eV (8.28(17) eV [2]), and $B$(M1;8.28 eV→0 eV) is the reduced transition probability for the M1 8.28 eV→0 eV transition in $\mu_N^2$. Using the Alaga rule, the $B$(M1;8.28 eV→0 eV) value can be obtained from the $B$(M1) values of the inter-band transitions between the 5/2$^+$[633] and 3/2$^+$[631] bands.

First, we derived the $B(X\lambda)$ values for the transitions related to the 5/2$^+$[633] and 3/2$^+$[631] bands, which are summarized in TABLE I. Here, the $B(X\lambda)$ values were calculated from the obtained half-lives, experimental γ branching ratios [1,11,17,22,33], theoretical mixing ratios [17] (partly experimental values [26,27]), and theoretical IC coefficients [31,36]. It is noteworthy that the γ branching ratios for some of the transitions starting with the 97.14-keV state were inconsistent among three measurements [11,17,22], and thus three $B(X\lambda)$ are listed in TABLE I for each transition starting with the 97.14-keV state.

Next, we investigated the validity of the Alaga rule for intra-band transitions in the 5/2$^+$[633] band. In TABLE II, the ratio of $B(X\lambda)$ of each intra-band transition in the 5/2$^+$[633] band to the value of 42.43 keV→0 keV is compared with the value estimated from the Alaga rule. For the 97.14 keV→42.43 keV and 97.14 keV→0 keV transitions, the $B$(M1) and $B$(E2) ratios obtained from the γ branching ratios in [17,22] (the values with superscripts a and b in TABLE II) were consistent with the values obtained from the Alaga rule within 1σ errors. This implies that the concept of separable rotational motion of axially deformed nuclei is applicable at these spin levels. By contrast, for the 163.25 keV→97.14 keV and 163.25 keV→42.43 keV transitions, we found that the $B$(M1) and $B$(E2) ratios deviated from the Alaga rule, as it often appears with increasing nuclear spins [37]. Furthermore, we investigated the validity of the Alaga rule for the intra-band transitions in the 3/2$^+$[631] band. In TABLE II, the ratio of each $B(X\lambda)$ to the value of the 29.19 keV→0.008 keV transition is compared with the value calculated from the Alaga rule. For the 71.83 keV→29.19 keV transition, the $B$(M1) and $B$(E2) ratios from the experimental values were consistent with those from the Alaga rule within 1σ errors. Hence, we conclude that the Alaga rule is valid for the intra-band transitions starting with the 29.19-, 42.43-, 71.83-, and 97.14-keV states.

Finally, we investigated whether the Alaga rule is applicable for inter-band transitions starting with the abovementioned four states to estimate $B$(M1;8.28 eV→0 eV). TABLE III shows the $B$(M1;8.28 eV→0 eV) values calculated from the $B$(M1) values for the inter-band transitions based on the Alaga rule. The $B$(M1;8.28 eV→0 eV) values from the 29.19 keV→0 keV and 71.83 keV→42.43 keV transitions agreed with each other. This indicates that the contribution of the Coriolis interactions, leading to band mixing and a deviation from the Alaga rule, is slight [10,13], and that the Alaga rule is applicable for the inter-band transitions between those low-spin states. The weighted average of $B$(M1;8.28 eV→0 eV) obtained from these transitions was 0.014(3) $\mu_N^2$, from which we estimated the radiative half-life of $^{229m}$Th to be 5.0(11)×10$^3$ s. Note that the contribution of the E2 transition to the radiative half-life is negligibly small [13].

A recent theoretical calculation based on the model of the collective quadrupole–octupole vibration-rotation motion of the nucleus including Coriolis interactions [12,38] indicated $B$(M1;8.28 eV→0 eV) values (0.011–0.014 $\mu_N^2$ [12]) similar to that estimated in this study. A more detailed comparison between theoretical and experimental results is necessitated to elucidate the nuclear structure of $^{229}$Th.

Meanwhile, the $B$(M1;8.28 eV→0 eV) values calculated from the $B$(M1) values of the 97.14 keV→71.83 keV transition (TABLE III) differed from our result of 0.014 $\mu_N^2$ (>3σ). As mentioned above, the $B$(M1) values of the 97.14 keV→71.83 keV transition are inconsistent among previous studies [11,17,22,30], primarily due to the difference in the γ branching ratios [11,17,22]. Therefore, further investigations of the 97.14 keV→71.83 keV transition are required to obtain consistent values.

## V. Conclusion and outlook

We performed coincidence measurements between α particles and γ rays emitted from a $^{233}$U source to determine the half-lives of the excited states in the $^{229}$Th nucleus. The half-lives of 42.43- and 164.53-keV states were determined to be 169(4) and 70(50) ps, respectively, which were consistent with literature values [11,32]. The half-life of the 97.14-keV state was determined to be 93(7) ps, which deviated from 147(12) ps reported in [32]. For the first time, the half-lives of 71.83- and 163.25-keV states were determined to be 120(40) and 220(30) ps, respectively. The $B$($X\lambda$) values related to the 5/2$^+$[633] and 3/2$^+$[631] bands, updated or newly determined in this study, were discussed; it was found that the Alaga rule is valid for the intra-band transitions except for the transitions starting with the 163.25-keV state. For the inter-band transitions between the 5/2$^+$[633] and 3/2$^+$[631] bands, we found that the $B$(M1) values for the 29.19 keV→0 keV and 71.83 keV→42.43 keV transitions reflected the Alaga rule. We estimated the $B$(M1) value for the 8.28 eV→0 eV transition to be 0.014(3) $\mu_N^2$, and the radiative half-life for $^{229m}$Th to be 5.0(11)×10$^3$ s.

The estimated half-life of $^{229m}$Th corresponds to the relative natural linewidth of the transition between the ground state and $^{229m}$Th of 2×10$^{-20}$, indicating the superb performance of the optical nuclear clock based on $^{229}$Th. Moreover, the obtained half-life facilitates the design of the experimental setup for the direct observation of radiation from $^{229m}$Th, leading to a more accurate determination of the radiative half-life. For example, experiments for directly observing the radiative decay of $^{229m}$Th were performed or are in progress by populating the isomeric state with synchrotron radiation [33,39–42]. In previous experiments [39–41], no photons originating from the decay of $^{229m}$Th were observed. The radiative half-life of $5.0(11)\times10^3$ s estimated in this study is outside of the scope of previous experiments [39–41]. Optimizing experimental conditions such as the irradiation time of synchrotron radiation may enable direct observation of radiation from $^{229m}$Th.

**Acknowledgements**

We are grateful to K. Yoshida and H. Watanabe for fruitful theoretical discussions. A. Yamaguchi acknowledges funding from the Technology Pioneering Projects by RIKEN. This study was supported by a JSPS Grant-in-Aid for Scientific Research (B) Grant No. JP18H01241, JP15H03661, and JP18H01230. The $^{233}$U sample used in this study was provided by the $^{233}$U cooperation project between Japan Atomic Energy Agency and the Inter-University Cooperative Research Program of the Institute for Materials Research, Tohoku University (Proposal No. 17K0204, 17F0011, and 18F0014).

**References**

[1]  B. R. Beck, J. A. Becker, P. Beiersdorfer, G. V. Brown, K. J. Moody, J. B. Wilhelmy, F. S. Porter, C. A. Kilbourne, and R. L. Kelley, *Energy Splitting of the Ground-State Doublet in the Nucleus $^{229}$Th*, Phys. Rev. Lett. **98**, 142501 (2007); Improved value for the energy splitting of the ground-state doublet in the nucleus $^{229m}$Th, Report No. LLNL-PROC-415170, 2009.

[2] B. Seiferle, L. von der Wense, P. V. Bilous, I. Amersdorffer, C. Lemell, F. Libisch, S. Stellmer, T. Schumm, C. E. Düllmann, A. Pálffy, and P. G. Thirolf, *Energy of the $^{229}$Th Nuclear Clock Transition*, Nature **573**, 243 (2019).

[3] A. Yamaguchi, H. Muramatsu, T. Hayashi, N. Yuasa, K. Nakamura, M. Takimoto, H. Haba, K. Konashi, M. Watanabe, H. Kikunaga, K. Maehata, N. Y. Yamasaki, and K. Mitsuda, *Energy of the $^{229}$Th Nuclear Clock Isomer Determined by Absolute γ-Ray Energy Difference*, Phys. Rev. Lett. **123**, 222501 (2019).

[4] T. Sikorsky, J. Geist, D. Hengstler, S. Kempf, L. Gastaldo, C. Enss, C. Mokry, J. Runke, C. E. Düllmann, P. Wobrauschek, K. Beeks, V. Rosecker, J. H. Sterba, G. Kazakov, T. Schumm, and A. Fleischmann, *Measurement of the $^{229}$Th Isomer Energy with a Magnetic Microcalorimeter*, Phys. Rev. Lett. **125**, 142503 (2020).

[5] E. Peik and C. Tamm, *Nuclear Laser Spectroscopy of the 3.5 eV Transition in Th-229*, Europhys. Lett. **61**, 181 (2003).

[6] C. J. Campbell, A. G. Radnaev, A. Kuzmich, V. A. Dzuba, V. V. Flambaum, and A. Derevianko, *Single-Ion Nuclear Clock for Metrology at the 19th Decimal Place*, Phys. Rev. Lett. **108**, 120802 (2012).

[7] L. von der Wense, B. Seiferle, M. Laatiaoui, J. B. Neumayr, H.-J. Maier, H.-F. Wirth, C. Mokry, J. Runke, K. Eberhardt, C. E. Düllmann, N. G. Trautmann, and P. G. Thirolf, *Direct Detection of the $^{229}$Th Nuclear Clock Transition*, Nature **533**, 47 (2016).

[8] B. Seiferle, L. von der Wense, and P. G. Thirolf, *Lifetime Measurement of the $^{229}$Th Nuclear Isomer*, Phys. Rev. Lett. **118**, 042501 (2017).

[9] Y. Shigekawa, Y. Kasamatsu, E. Watanabe, H. Ninomiya, S. Hayami, N. Kondo, Y. Yasuda, H. Haba, and A. Shinohara, *Observation of Internal-Conversion Electrons Emitted from $^{229m}$Th Produced by β Decay of $^{229}$Ac*, Phys. Rev. C **100**, 044304 (2019).

[10] A. M. Dykhne and E. V. Tkalya, *Matrix Element of the Anomalously Low-Energy (3.5±0.5 eV) Transition in $^{229}$Th and the Isomer Lifetime*, J. Exp. Theor. Phys. Lett. **67**, 251 (1998).

[11] E. Ruchowska, W. A. Płóciennik, J. Żylicz, H. Mach, J. Kvasil, A. Algora, N. Amzal, T. Bäck, M. G. Borge, R. Boutami, P. A. Butler, J. Cederkäll, B. Cederwall, B. Fogelberg, L. M. Fraile, H. O. U. Fynbo, E. Hagebø, P. Hoff, H. Gausemel, A. Jungclaus, R. Kaczarowski, A. Kerek, W. Kurcewicz, K. Lagergren, E. Nacher, B. Rubio, A. Syntfeld, O. Tengblad, A. A. Wasilewski, and L. Weissman, *Nuclear Structure of $^{229}$Th*, Phys. Rev. C **73**, 044326 (2006).

[12] N. Minkov and A. Pálffy, *Reduced Transition Probabilities for the Gamma Decay of the 7.8 eV Isomer in $^{229}$Th*, Phys. Rev. Lett. **118**, 212501 (2017).

[13] E. V. Tkalya, C. Schneider, J. Jeet, and E. R. Hudson, *Radiative Lifetime and Energy of the Low-Energy Isomeric Level in $^{229}$Th*, Phys. Rev. C **92**, 054324 (2015).

[14] G. Alaga, K. Alder, A. Bohr, and B. R. Mottelson, *INTENSITY RULES FOR BETA AND GAMMA TRANSITIONS TO NUCLEAR ROTATIONAL STATES*, Dan Mat Fys Medd **29** (No.9), 1 (1955).

[15] L. A. Kroger and C. W. Reich, *Features of the Low-Energy Level Scheme of $^{229}$Th as Observed in the α-Decay of $^{233}$U*, Nucl. Phys. A **259**, 29 (1976).

[16] M. J. Canty, R. D. Connor, D. A. Dohan, and B. Pople, *The Decay of $^{233}$U*, J. Phys. G **3**, 421 (1977).

[17] V. Barci, G. Ardisson, G. Barci-Funel, B. Weiss, O. El Samad, and R. K. Sheline, *Nuclear Structure of $^{229}$Th from γ-Ray Spectroscopy Study of $^{233}$U α-Particle Decay*, Phys. Rev. C **68**, 034329 (2003).

[18] C. W. Reich, R. G. Helmer, J. D. Baker, and R. J. Gehrke, *Emission Probabilities and Energies of γ-Ray Transitions from the Decay of $^{233}$U*, Int. J. Appl. Radiat. Isot. **35**, 185 (1984).

[19] C. W. Reich and R. G. Helmer, *Energy Separation of the Doublet of Intrinsic States at the Ground State of $^{229}$Th*, Phys. Rev. Lett. **64**, 271 (1990).

[20] R. G. Helmer and C. W. Reich, *An Excited State of $^{229}$Th at 3.5 eV*, Phys. Rev. C **49**, 1845 (1994).

[21] K. Chayawattanangkur, G. Herrmann, and N. Trautmann, *Heavy Isotopes of Actinium: $^{229}Ac$, $^{230}Ac$, $^{231}Ac$ and $^{232}Ac$*, J. Inorg. Nucl. Chem. **35**, 3061 (1973).

[22] K. Gulda, W. Kurcewicz, A. J. Aas, M. J. G. Borge, D. G. Burke, B. Fogelberg, I. S. Grant, E. Hagebø, N. Kaffrell, J. Kvasil, G. Løvhøiden, H. Mach, A. Mackova, T. Martinez, G. Nyman, B. Rubio, J. L. Tain, O. Tengblad, and T. F. Thorsteinsen, *The Nuclear Structure of $^{229}Th$*, Nucl. Phys. A **703**, 45 (2002).

[23] I. Ahmad, J. E. Gindler, R. R. Betts, R. R. Chasman, and A. M. Friedman, *Possible Ground-State Octupole Deformation in $^{229}Pa$*, Phys. Rev. Lett. **49**, 1758 (1982).

[24] K. M. Glover, *Alpha-Particle Spectrometry and Its Applications*, Int. J. Appl. Radiat. Isot. **35**, 239 (1984).

[25] I. Ahmad, *Relative Alpha Intensities of Several Actinide Nuclides*, Nucl. Instrum. Methods Phys. Res. **223**, 319 (1984).

[26] E. F. Tretyakov, M. P. Anikina, L. L. Goldin, G. I. Novikova, and N. I. Pirogova, *THE SPECTRUM OF INTERNAL CONVERSION ELECTRONS ACCOMPANYING ALPHA DECAY OF $U^{233}$, AND THE LEVEL SCHEME OF $Th^{229}$*, Zh. Eksp. Teor. Fiz. **37**, 917 (1959) [Sov. Phys. JETP **10**, 656 (1960)].

[27] T. Andersen, K. M. Bisgard, and P. G. Hansen, *Low Lying States of $Th^{229}$*, Nucl. Phys. **27**, 73 (1961).

[28] D. G. Burke, P. E. Garrett, T. Qu, and R. A. Naumann, *Additional Evidence for the Proposed Excited State at ≤5 eV in $^{229}Th$*, Phys. Rev. C **42**, R499 (1990).

[29] D. G. Burke, P. E. Garrett, T. Qu, and R. A. Naumann, *Nuclear Structure of $^{229,231}Th$ Studied with the $^{230,232}Th(d,t)$ Reactions*, Nucl. Phys. A **809**, 129 (2008).

[30] C. E. Bemis, F. K. McGowan, J. L. C. Ford, W. T. Milner, R. L. Robinson, and P. H. Stelson, *Coulomb Excitation of States in $^{229}Th$*, Phys. Scr. **38**, 657 (1988).

[31] E. Browne and J. K. Tuli, *Nuclear Data Sheets for A = 229*, Nucl. Data Sheets **109**, 2657 (2008).

[32] H. Ton, S. Roodbergen, J. Brasz, and J. Blok, *LIFETIMES OF ROTATIONAL STATES IN $^{229}$Th AND $^{235}$U*, Nucl. Phys. A **155**, 245 (1970).

[33] T. Masuda, A. Yoshimi, A. Fujieda, H. Fujimoto, H. Haba, H. Hara, T. Hiraki, H. Kaino, Y. Kasamatsu, S. Kitao, K. Konashi, Y. Miyamoto, K. Okai, S. Okubo, N. Sasao, M. Seto, T. Schumm, Y. Shigekawa, K. Suzuki, S. Stellmer, K. Tamasaku, S. Uetake, M. Watanabe, T. Watanabe, Y. Yasuda, A. Yamaguchi, Y. Yoda, T. Yokokita, M. Yoshimura, and K. Yoshimura, *X-ray pumping of the $^{229}$Th nuclear clock isomer*, Nature **573**, 238 (2019).

[34] H. Kikunaga, T. Suzuki, M. Nomura, T. Mitsugashira, and A. Shinohara, *Determination of the Half-Life of the Ground State of $^{229}$Th by Using $^{232}$U and $^{233}$U Decay Series*, Phys. Rev. C **84**, 014316 (2011).

[35] N. Trautmann and H. Folger, *Preparation of Actinide Targets by Electrodeposition*, Nucl. Instrum. Methods A **282**, 102 (1989).

[36] T. Kibédi, T. W. Burrows, M. B. Trzhaskovskaya, P. M. Davidson, and C. W. Nestor, *Evaluation of Theoretical Conversion Coefficients Using BrIcc*, Nucl. Instrum. Methods A **589**, 202 (2008).

[37] R. F. Casten, *Nuclear Structure from a Simple Perspective* (Oxford University Press, New York, 1990).

[38] N. Minkov and A. Pálffy, *$^{229m}$Th Isomer from a Nuclear Model Perspective*, Phys. Rev. C **103**, 014313 (2021).

[39] J. Jeet, C. Schneider, S. T. Sullivan, W. G. Rellergert, S. Mirzadeh, A. Cassanho, H. P. Jenssen, E. V. Tkalya, and E. R. Hudson, *Results of a Direct Search Using Synchrotron Radiation for the Low-Energy $^{229}$Th Nuclear Isomeric Transition*, Phys. Rev. Lett. **114**, 253001 (2015).

[40] A. Yamaguchi, M. Kolbe, H. Kaser, T. Reichel, A. Gottwald, and E. Peik, *Experimental Search for the Low-Energy Nuclear Transition in $^{229}$Th with Undulator Radiation*, New J. Phys. **17**, 053053 (2015).

[41] S. Stellmer, G. Kazakov, M. Schreitl, H. Kaser, M. Kolbe, and T. Schumm, *Attempt to Optically Excite the Nuclear Isomer in $^{229}$Th*, Phys. Rev. A **97**, 062506 (2018).

[42] S. Stellmer, M. Schreitl, G. Kazakov, K. Yoshimura, and T. Schumm, *Towards a Measurement of the Nuclear Clock Transition in $^{229}$Th*, J. Phys. Conf. Ser. **723**, 012059 (2016).

**Figure**

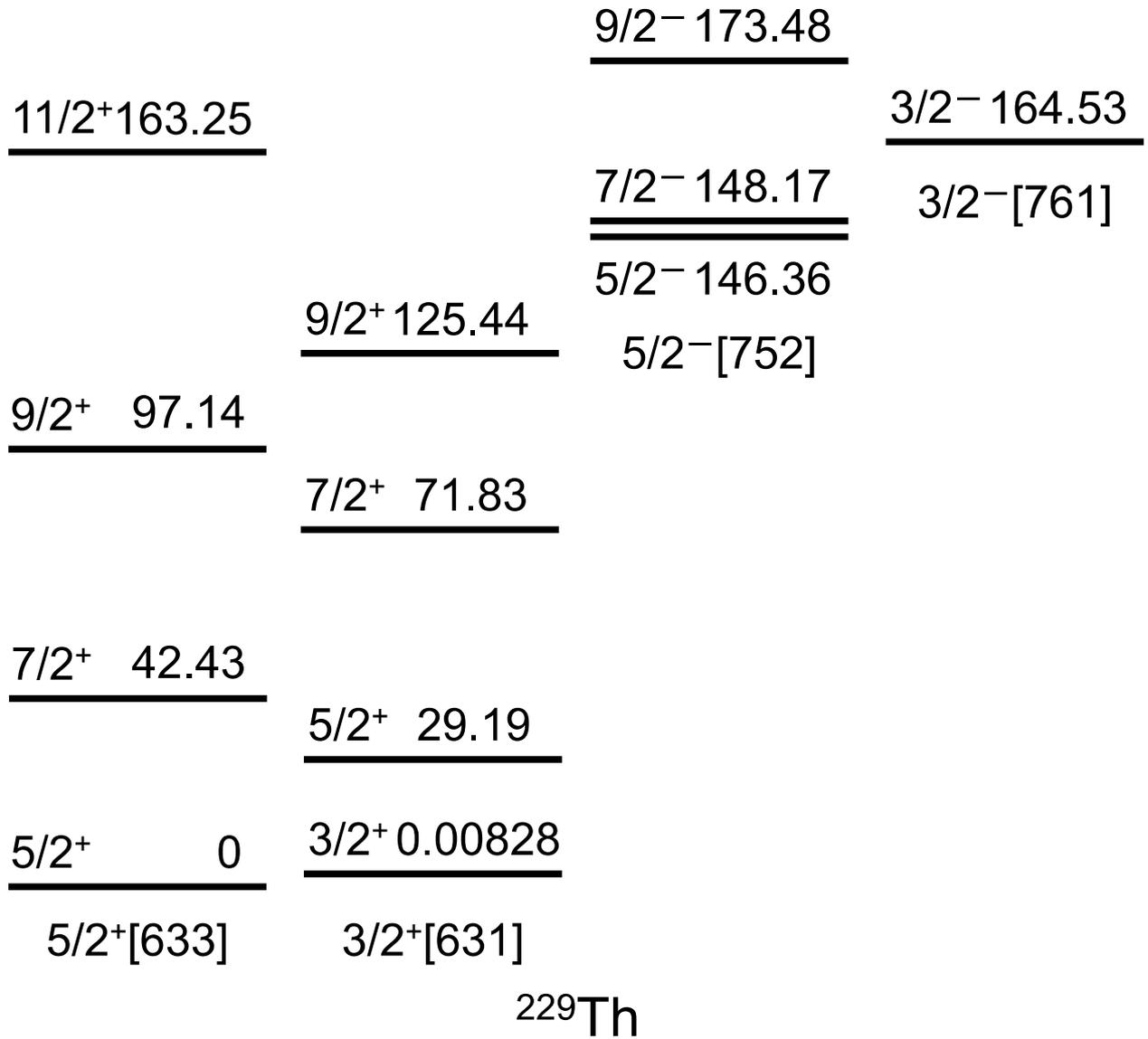

Fig. 1. Energy levels of the $^{229}$Th nucleus grouped into rotational bands [31], which are related to this study (units: keV). The energy of $^{229m}$Th is derived from [2].

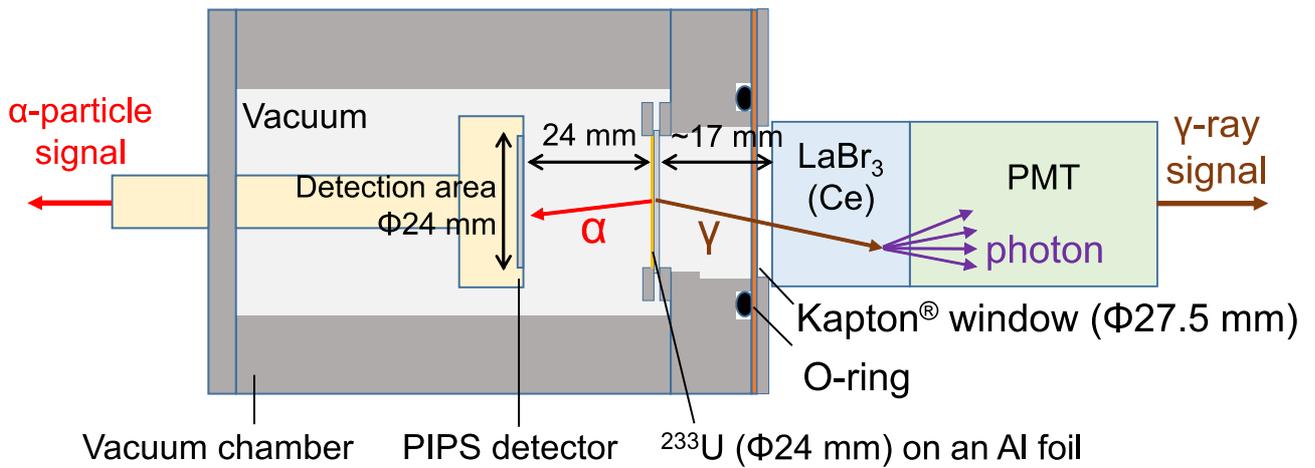

Fig. 2. Setup for the coincidence measurement between α particles and γ rays from a $^{233}$U source.

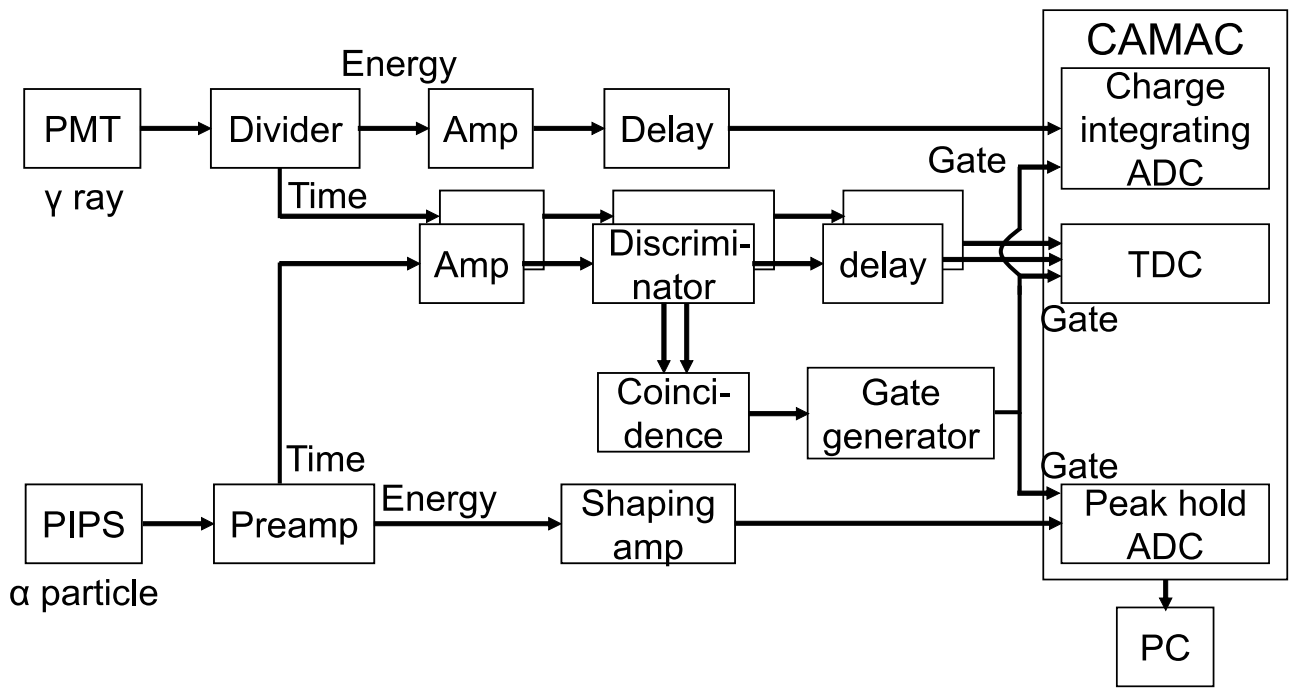

Fig. 3. Block diagram of the coincidence measurement.

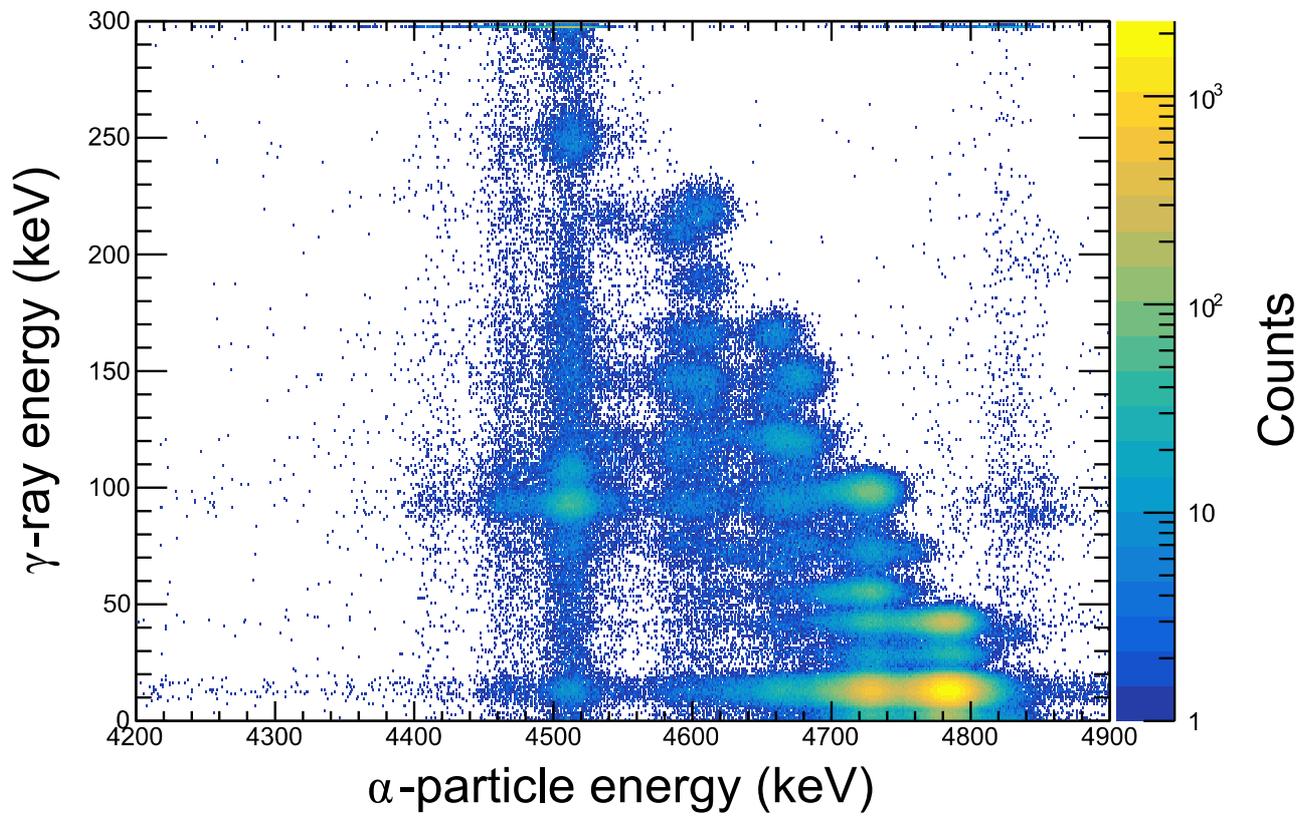

Fig. 4. 2D energy spectrum of coincided α-particle and γ-ray events.

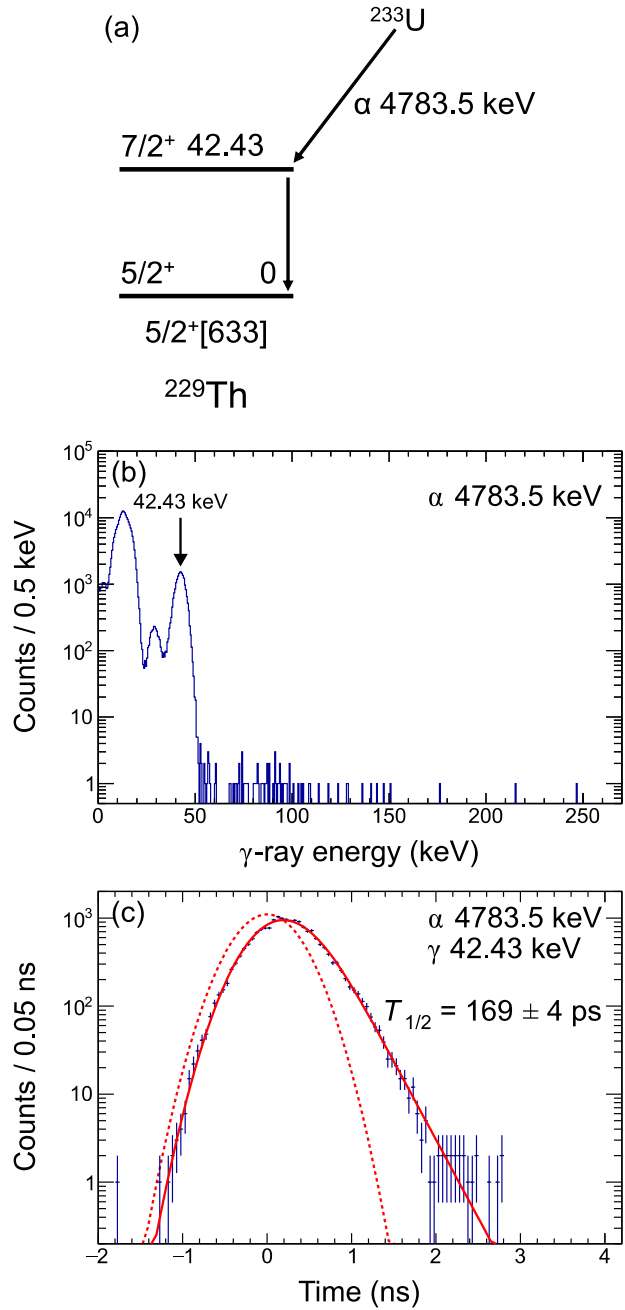

Fig. 5. (a) Energy levels and transitions relevant to the half-life determination of the 42.43-keV state (units: keV). (b) γ-ray spectrum gated by the 4783.5-keV α peak. The arrow shows the 42.43-keV γ peak. (c) Time trace of the 42.43-keV γ-ray signals following the 4783.5-keV α-particle signals. The error bars indicate the statistical uncertainty of 1σ. The solid red curve shows the function fitted to the data, and the dashed red curve shows the Gaussian component in the fitted function. The FWHM of the Gaussian component was determined to be 821(8) ps.

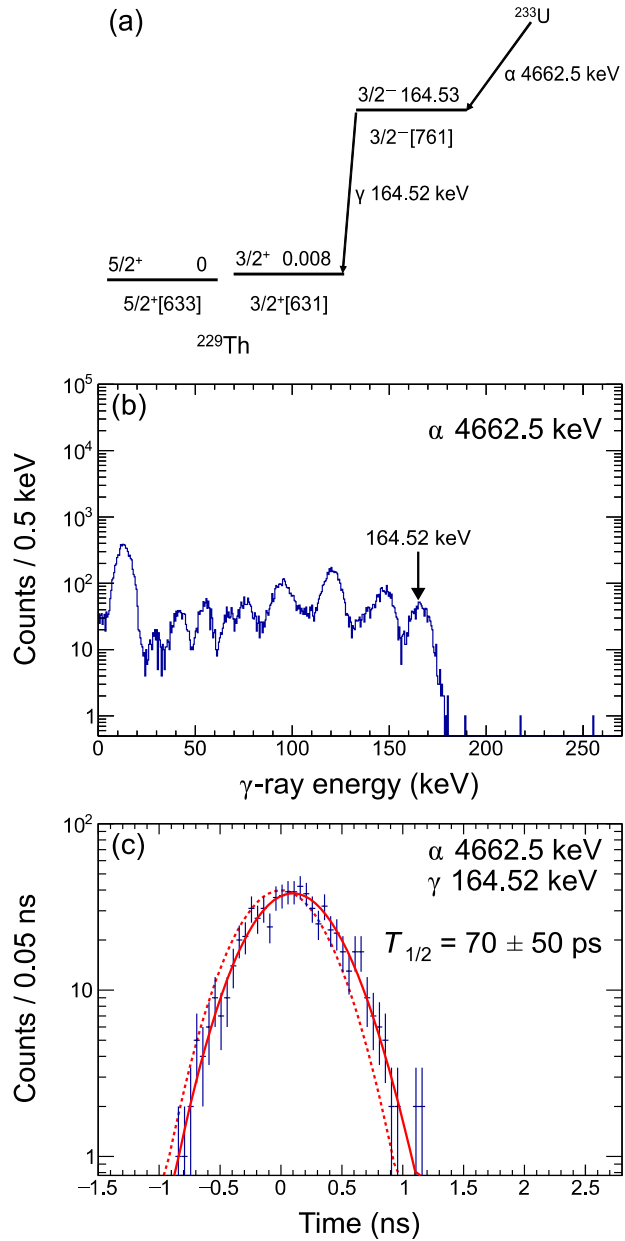

Fig. 6. (a) Energy levels and transitions, relevant to the half-life determination of the 164.53-keV state (units: keV). (b) γ-ray spectrum gated by the 4662.5-keV α peak. The arrow shows the 164.52-keV γ peak. (c) Time trace of the 164.52-keV γ-ray signals following the 4662.5-keV α-particle signals. The error bars indicate the statistical uncertainty of 1σ. The solid red curve shows the function fitted to the data, and the dashed red curve shows the Gaussian component in the fitted function. The FWHM of the Gaussian component was determined to be 800(50) ps.

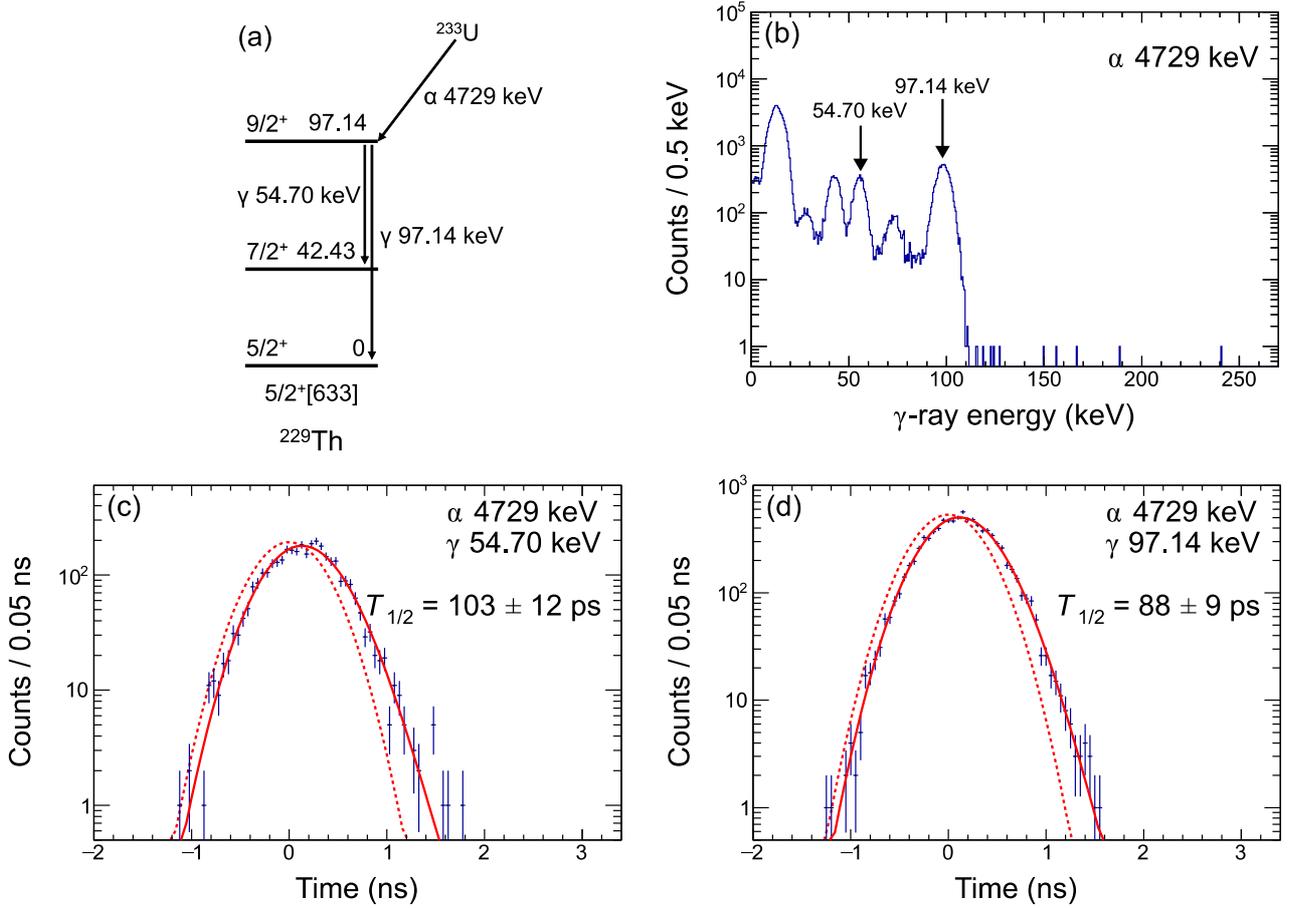

Fig. 7. (a) Energy levels and transitions, relevant to the half-life determination of the 97.14-keV state (units: keV). (b) γ-ray spectrum gated by the 4729-keV α peak. The arrows on the left and right sides show the 54.70- and 97.14-keV γ peaks, respectively. (c) Time trace of the 54.70-keV γ-ray signals following the 4729-keV α-particle signals. (d) Time trace of the 97.14-keV γ-ray signals following the 4729-keV α-particle signals. The error bars in (c) and (d) indicate the statistical uncertainty of 1σ. The solid red curves show the functions fitted to the data, and the dashed red curves show the Gaussian components in the fitted functions. The FWHMs of the Gaussian components in (c) and (d) were determined to be 807(19) and 794(12) ps, respectively.

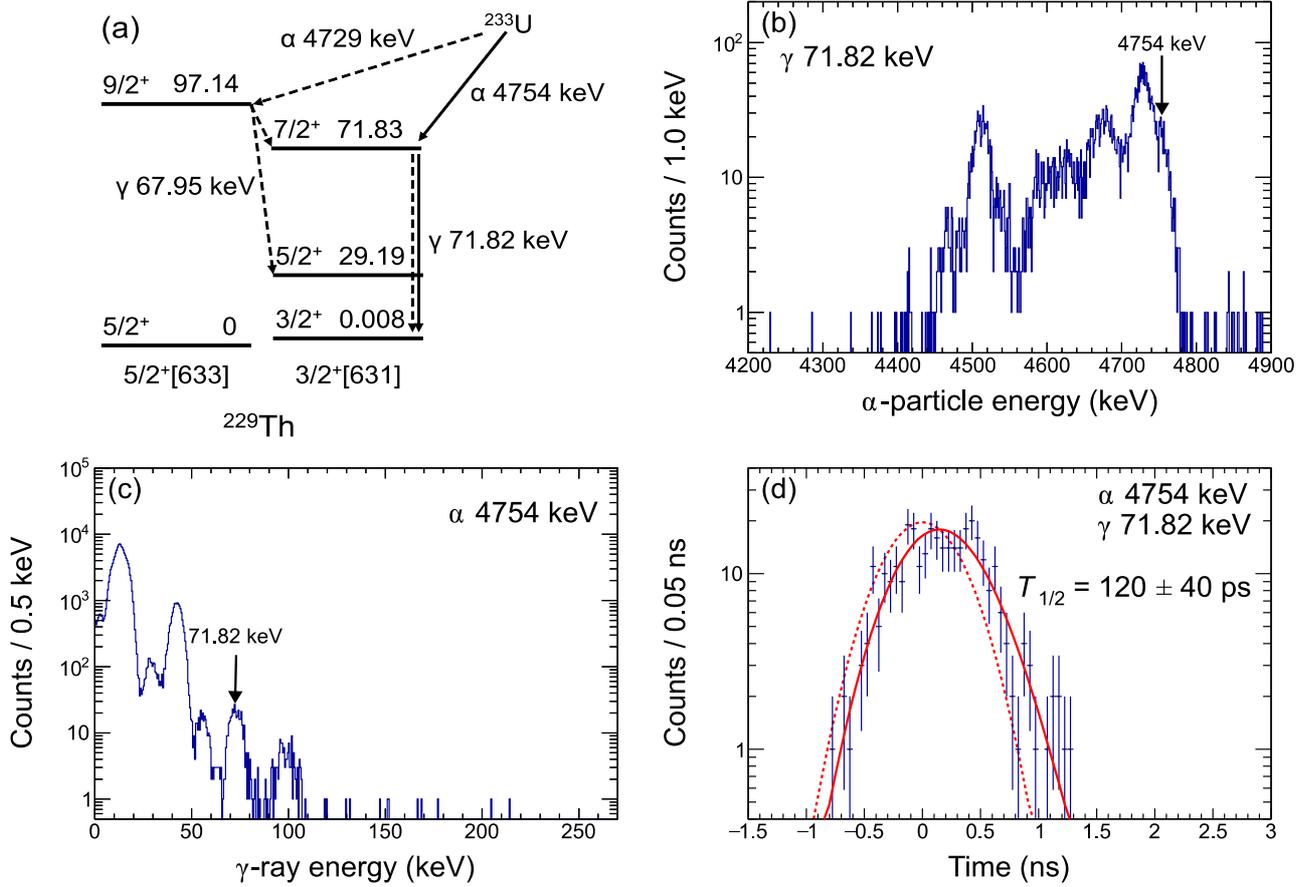

Fig. 8. (a) Energy levels and transitions, relevant to the half-life determination of the 71.83-keV state (units: keV). The dashed lines represent the transitions that may interfere with the half-life determination. (b) α-particle spectrum gated by the 71.82-keV γ peak. The arrow shows the 4754-keV α peak. (c) γ-ray spectrum gated by the 4754-keV α peak. The arrow shows the 71.82-keV γ peak. (d) Time trace of the 71.82-keV γ-ray signals following the 4754-keV α-particle signals. The error bars indicate the statistical uncertainty of 1σ. The solid red curve shows the function fitted to the data, and the dashed red curve shows the Gaussian component in the fitted function. The FWHM of the Gaussian component was determined to be 790(60) ps.

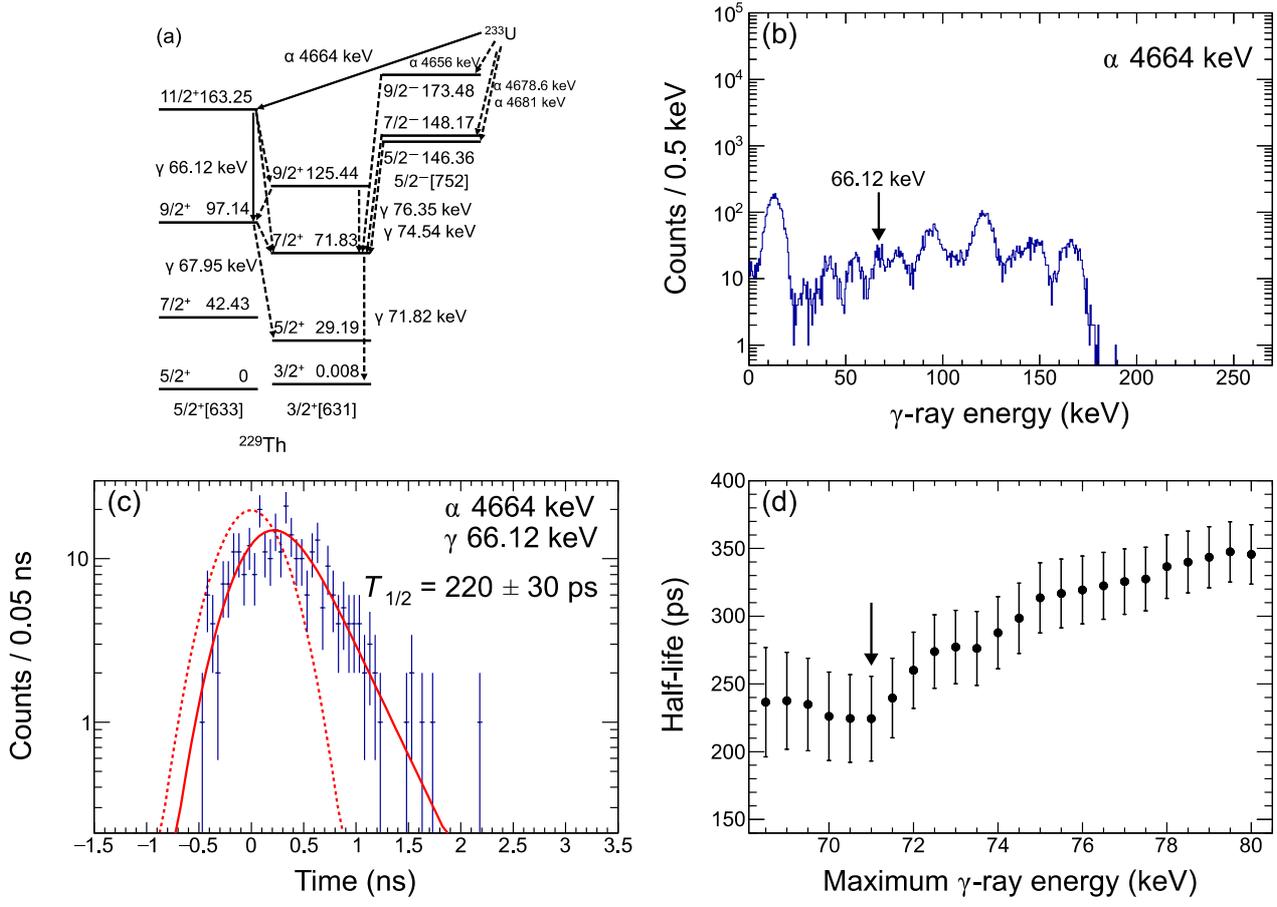

Fig. 9. (a) Energy levels and transitions, relevant to the half-life determination of the 163.25-keV state (units: keV). The dashed lines represent the transitions that may interfere with the half-life determination. (b) γ-ray spectrum gated by the 4664-keV α peak. The arrow shows the 66.12-keV γ peak. (c) Time trace of the 66.12-keV γ-ray signals following the 4664-keV α-particle signals. The error bars indicate the statistical uncertainty of 1σ. The solid red curve shows the function fitted to the data, and the dashed red curve shows the Gaussian component in the fitted function. (d) Half-lives obtained by the fitting as a function of the maximum energy of the γ energy window for the time traces of the 66.12-keV γ rays following the 4664-keV α particles. The minimum energy of the γ energy window was fixed to 64 keV. The FWHM of the Gaussian component was determined to be 670(70) ps.

Tables

TABLE I. Reduced transition probabilities $B(X\lambda)$ newly determined or upgraded in this study.

| $I_i^\pi K_i$ | Initial level (keV) | $T_{1/2}$ (ps) | $I_f^\pi K_f$ | Final level (keV) | $b_\gamma$ (%)[a] | $\delta$ [b] | $\alpha$ [c] | $X\lambda$ | $B(X\lambda)$[d] |
|---|---|---|---|---|---|---|---|---|---|
| 5/2+3/2 | 29.19 | 82.2(40)[e] | 3/2+3/2 | 0.008 | 89.4(3)[e] | 0.145(10) | 225(13) | M1 | 0.049(9) |
| | | | | | | | | E2 | 1.7(4) |
| | | | 5/2+5/2 | 0 | 10.6(3)[e] | 0.65(16)[f] | 1370(410)[e] | M1 | 0.0041(10) |
| | | | | | | | | E2 | 3(1) |
| 7/2+5/2 | 42.43 | 169(4) | 5/2+3/2 | 29.19 | 3.2(9)[g] | 0.0030(3) | 359(5) | M1 | 0.023(9) |
| | | | | | | | | E2 | 0.0017(7) |
| | | | 3/2+3/2 | 0.008 | 0.24(7)[g] | – | 684(10) | E2 | 0.042(17) |
| | | | 5/2+5/2 | 0 | 96.5(9)[h] | 0.40(10)[i] | 130(40) | M1 | 0.018(5) |
| | | | | | | | | E2 | 2.3(12) |
| 7/2+3/2 | 71.83 | 120(40) | 7/2+5/2 | 42.43 | 2.2(3)[j] | 0.041(4) | 143.7(25) | M1 | 0.0049(17) |
| | | | | | | | | E2 | 0.014(5) |
| | | | 5/2+3/2 | 29.19 | 79.9(12)[h] | 0.143(12) | 58.2(23) | M1 | 0.059(19) |
| | | | | | | | | E2 | 1.0(3) |
| | | | 3/2+3/2 | 0.008 | 11.0(9)[h] | – | 53.8(8) | E2 | 0.48(16) |
| | | | 5/2+5/2 | 0 | 7.0(7)[g] | 0.248(22) | 12.5(5) | M1 | 0.0010(4) |
| | | | | | | | | E2 | 0.018(7) |
| 9/2+5/2 | 97.14 | 93(7) | 7/2+3/2 | 71.83 | 5.3(3)[h] | 0.0044(4) | 213(3) | M1 | 0.033(3) |
| | | | | | 2.8(5)[k] | | | | 0.019(3) |
| | | | | | 13.2(17)[l] | | | | 0.059(7) |
| | | | | | | | | E2 | 0.0014(3) |
| | | | | | | | | | 0.00083(21) |
| | | | | | | | | | 0.0026(6) |
| | | | 7/2+5/2 | 42.43 | 42.5(16)[h] | 0.46(3)[i] | 53(4) | M1 | 0.021(2) |
| | | | | | 45.5(36)[k] | | | | 0.025(3) |
| | | | | | 39.4(56)[l] | | | | 0.014(2) |
| | | | | | | | | E2 | 2.2(3) |
| | | | | | | | | | 2.6(4) |
| | | | | | | | | | 1.5(3) |
| | | | 5/2+3/2 | 29.19 | 0.81(6)[h] | | 70.2(10) | E2 | 0.081(10) |
| | | | | | 0.73(15)[k] | | | | 0.080(18) |
| | | | | | 3.4(9)[l] | | | | 0.25(7) |
| | | | 5/2+5/2 | 0 | 51.4(16)[h] | | 12.91(18) | E2 | 0.86(8) |
| | | | | | 51.0(38)[k] | | | | 0.94(12) |
| | | | | | 44.0(63)[l] | | | | 0.53(9) |
| 11/2+5/2 | 163.25 | 220(30) | 9/2+3/2 | 125.44 | 6.0(9)[h] | 0.0162(14) | 65.4(10) | M1 | 0.012(3) |
| | | | | | | | | E2 | 0.0033(9) |
| | | | 9/2+5/2 | 97.14 | 25.4(14)[h] | 0.56(20) | 29(9) | M1 | 0.0074(20) |
| | | | | | | | | E2 | 0.8(4) |
| | | | 7/2+3/2 | 71.83 | 1.0(2)[g] | | 17.15(24) | E2 | 0.025(7) |
| | | | 7/2+5/2 | 42.43 | 67.6(17)[h] | | 4.95(7) | E2 | 0.42(9) |

[a] Branching ratio of a γ ray from an initial level.

[b] Theoretically calculated mixing ratio [17] unless otherwise stated.

[c] Internal conversion coefficient from [31,36] unless otherwise stated.

[d] Reduced transition probabilities in $\mu_N^2$ for M1 and $e^2b^2$ for E2.

[e] Taken from experimentally determined values in [33].

[f] Calculated from $\alpha$ = 1370(410) [33,36].

[g] Taken from the γ-ray intensity calculated from strong rotational model [17]. As described in [17], α-particle intensities obtained from the calculated γ-ray intensities agree well with experimental α-particle measurements.

[h] Taken from the experimentally determined γ-ray intensity [17].

[i] Experimentally determined from the spectroscopy of internal conversion electrons [26,27,31].

[j] Taken from [1]. Branching ratio for the 71.83 keV→42.43 keV transition is 1/37 of that for the 71.83 keV→29.19 keV transition.

[k] Taken from the experimentally determined γ-ray intensity [22].

[l] Taken from the experimentally determined γ-ray intensity [11].

TABLE II. Comparison of the ratio of the experimentally obtained $B(X\lambda)$ with the value estimated from the Alaga rule for intra-band transitions in the $5/2^+[633]$ and $3/2^+[631]$ bands.

| Band | $X\lambda$ | $I_i^\pi$ | Initial level (keV) | $I_f^\pi$ | Final level (keV) | $B(X\lambda)$ ratio Exp. | Alaga |
|---|---|---|---|---|---|---|---|
| 5/2$^+$[633] | M1 | 7/2$^+$ | 42.43 | 5/2$^+$ | 0 | 1 | 1 |
| | | 9/2$^+$ | 97.14 | 7/2$^+$ | 42.43 | 1.2(4)[a] | 1.45 |
| | | | | | | 1.4(4)[b] | 1.45 |
| | | | | | | 0.8(3)[c] | 1.45 |
| | | 11/2$^+$ | 163.25 | 9/2$^+$ | 97.14 | 0.4(2) | 1.7 |
| | E2 | 7/2$^+$ | 42.43 | 5/2$^+$ | 0 | 1 | 1 |
| | | 9/2$^+$ | 97.14 | 7/2$^+$ | 42.43 | 0.9(5)[a] | 0.85 |
| | | | | | | 1.1(6)[b] | 0.85 |
| | | | | | | 0.6(3)[c] | 0.85 |
| | | | | 5/2$^+$ | 0 | 0.4(2)[a] | 0.28 |
| | | | | | | 0.4(2)[b] | 0.28 |
| | | | | | | 0.2(1)[c] | 0.28 |
| | | 11/2$^+$ | 163.25 | 9/2$^+$ | 97.14 | 0.3(3) | 0.66 |
| | | | | 7/2$^+$ | 42.43 | 0.2(1) | 0.48 |
| 3/2$^+$[631] | M1 | 5/2$^+$ | 29.19 | 3/2$^+$ | 0.008 | 1 | 1 |
| | | 7/2$^+$ | 71.83 | 5/2$^+$ | 29.19 | 1.2(4) | 1.34 |
| | E2 | 5/2$^+$ | 29.19 | 3/2$^+$ | 0.008 | 1 | 1 |
| | | 7/2$^+$ | 71.83 | 5/2$^+$ | 29.19 | 0.6(2) | 0.63 |
| | | | | 3/2$^+$ | 0.008 | 0.3(1) | 0.42 |

[a] Based on the γ-ray intensity ratios from [17].

[b] Based on the γ-ray intensity ratios from [22]

[c] Based on the γ-ray intensity ratios from [11]

TABLE III. $B$(M1) for the 8.28 eV→0 eV transition and the radiative half-life of $^{229m}$Th, calculated from the experimentally obtained $B$(M1) for the inter-band transitions between the 5/2$^+$[633] and 3/2$^+$[631] bands based on the Alaga rule. Theoretically calculated $B$(M1;8.28 eV→0 eV) [10–12] are also shown. The values originating from the 42.43 keV→29.19 keV transition are not listed because the γ branching ratio for this transition has not yet been measured experimentally. The $B$(M1;8.28 eV→0 eV) values obtained from the $B$(M1) values of the 97.14 keV→71.83 keV transition based on [11,17,22] differed from those listed in [13] mainly because the half-life of the 97.14-keV state was updated in this study.

| | Experiment | | | | |
|---|---|---|---|---|---|
| $I_i^\pi K_i$ | Initial level (keV) | $I_f^\pi K_f$ | Final level (keV) | $B$(M1;8.28 eV→0 eV) ($\mu_N^2$) | $T_{1/2}$ of $^{229m}$Th (s) |
| 5/2$^+$3/2 | 29.19 | 5/2$^+$5/2 | 0 | 0.014(4) | 4.8(12)×10$^3$ |
| 7/2$^+$3/2 | 71.83 | 7/2$^+$5/2 | 42.43 | 0.013(4) | 5.4(19)×10$^3$ |
| 9/2$^+$5/2 | 97.14 | 7/2$^+$3/2 | 71.83 | 0.071(7)[a] | 9.8(12)×10$^2$ |
| | | | | 0.041(7)[b] | 1.7(3)×10$^3$ |
| | | | | 0.13(1)[c] | 5.5(7)×10$^2$ |
| | | | | 0.069(13)[d] | 1.0(2)×10$^3$ |
| 11/2$^+$5/2 | 163.25 | 9/2$^+$3/2 | 125.44 | 0.029(7) | 2.4(6)×10$^3$ |
| Theory | | | | | |
| Ref. [10] | | | | 0.086 | 8.1(5)×10$^{2}$ [e] |
| Ref. [11] | | | | 0.025 | 2.8(2)×10$^{3}$ [e] |
| Ref. [12] | | | | 0.011–0.014 | 4.8(3)×10$^3$– 6.5(4)×10$^{3}$ [e] |

[a] Based on the γ-ray intensity ratios from [17].

[b] Based on the γ-ray intensity ratios from [22].

[c] Based on the γ-ray intensity ratios from [11]

[d] From $B$(M1;97.14 keV→71.83 keV) = 0.032(6) $\mu_N^2$, experimentally obtained in [30].

[e] The error comes from the error of the isomer energy [2].